# Verification and Diagnosis Infrastructure of SoC HDL-model

Vladimir Hahanov[1], Wajeb Gharibi[2], Eugenia Litvinova[1], Svetlana Chumachenko[1]
[1]Computer Engineering Faculty, Kharkov National University of Radioelectronics, Kharkov, Ukraine,
hahanov@kture.kharkov.ua
[2]Jazan University, Jazan, Kingdom of Saudi Arabia, Gharibi@jazanu.edu.sa

*Abstract* — **This article describes technology for diagnosing SoC HDL-models, based on transactional graph. Diagnosis method is focused to considerable decrease the time of fault detection and memory for storage of diagnosis matrix by means of forming ternary relations in the form of test, monitor, and functional component. The following problems are solved: creation of digital system model in the form of transaction graph and multi-tree of fault detection tables, as well as ternary matrices for activating functional components in tests, relative to the selected set of monitors; development of a method for analyzing the activation matrix to detect the faults with given depth and synthesizing logic functions for subsequent embedded hardware fault diagnosing.**

*Keywords-system-on-chip; diagnosis; fault detection table; transaction graph, SoC HDL- model; infrastructure IP.*

## I. TAB-MODEL FOR DIAGNOSIS FAULTY COMPONENTS OF SoC HDL-MODEL

Motivation of this paper is determined by the following: 1) the creation of simple and applicable models, methods and engines for diagnosis of multilayer software and hardware systems; 2) market appeal of matrix or table method for fault detection of SoC IP components (hardware and software) as the most effective one, which is focused on parallel processing and makes it possible to considerably reduce the time of faults diagnosis in non-functional mode.

Aim of this article is creation of model and method for considerable decrease the testing time and memory for storage of diagnosis matrix by means of forming ternary relations (test – monitor – IP-function) in a single table TAB: Tests – Assertions – Blocks. The problems and background as published papers are: 1) development of diagnosability criteria and HDL digital system model in the form of transaction graph, as well as multi-level model and engine for diagnosis of software and hardware modules, based on activation matrix of functional components by using tests responded on the selected monitor set [1-6]; 2) development of a fault diagnosis method based on activation matrix analysis with a given depth [4-7]; 3) Implementation of diagnosis infrasructure in the system Riviera, Aldec [8-11].

Model for diagnosis of HDL digital system model is represented by the following transformation of the initial diagnosis equation, defined by xor-relation of the parameters <test – functionality – faulty blocks>:

$$T \oplus F \oplus B = 0 \rightarrow B = T \oplus F \rightarrow B = \{T \times A\} \oplus F \rightarrow$$
$$\rightarrow B = \{T \times A\} \oplus \{F \times m\},$$

which is transformed in ternary matrix relation of the T,F,B:

$$M = \{\{T \times A\} \times \{B\}\} \leftarrow M_{ij} = (T \times A)_i \oplus B_j.$$

Here, the coordinate of matrix (table) is equal to 1, if the pair test–monitor (assertions) $(T \times A)_i$ activates faults of the functional block $B_j \in B$ on the monitors.

A digital system model is presented as transaction graph:

$$G = <B, A>, B = \{B_1, B_2, \ldots, B_i, \ldots, B_n\},$$
$$A = \{A_1, A_2, \ldots, A_j, \ldots, A_m\},$$

where sets of arcs $B$ – functional blocks; $A$ – monitors for observation of the HDL-code variables are defined.

Test segments set $T = \{T_1, T_2, \ldots, T_r, \ldots, T_k\}$ for HDL-blocks diagnosis overlaps the graph model, and activates the transaction paths in the graph. In general, the testing model is represented by the Cartesian product $M = <B \times A \times T>$ that has the dimension $Q = n \times m \times k$. To reduce the amount of diagnostic information it is offered to assign assertion monitor to each test segment, which answers for visualization of an activation way of functional blocks. It makes possible to decrease the dimension of model (matrix) to $Q = n \times k$ and retain all features of the triad relationship $M = <B \times A \times T>$. For the pair «test – monitor» not only one-to-one correspondence is possible $<T_i \rightarrow A_j>$, but functional $<\{T_i, T_r\} \rightarrow A_j>$ and injective ones $<T_i \rightarrow \{A_j, A_s\}>$ are used too. Such variety of correspondences makes it possible to duplicate one test segment for different monitors, as well as assign several tests to the same monitor. At that the matrix cell $M_{ij} = \{0,1\}$ always preserves its dimension, equal to 1 bit.

The analytical matrix diagnosis model by using monitor engine, focused to achieving a given depth of faulty code diagnosis, is presented in the following form:

$$M = f(G,L,T,B,A,t), B = \{B_1, B_2, \ldots, B_i, \ldots, B_m\};$$
$$L = \{L_1, L_2, \ldots, L_i, \ldots, L_n\}; A(t) = \{A_1, A_2, \ldots, A_i, \ldots, A_k\};$$
$$A \subseteq L; G = L \times B; k \leq n; T = \{T_1, T_2, \ldots, T_i, \ldots, T_p\}.$$

Here $B_i$ is a group of code statements, assigned to the node $L_i$ (variable, register, counter, memory), determining its state; G is functionality, presented by the transaction graph $G = (L, A) \times B$ in the form of the Cartesian product of node and arc sets; A is a set of monitors, as a subset of transaction graph nodes $A \subseteq L$. The method for detecting faults of the functional blocks (FB) uses pre-built activation table (matrix) ATFB $M = [M_{ij}]$, where row is the relation between the test segment and a subset of activated blocks

$$T_i \to A_j \approx (M_{i1}, M_{i2}, \ldots, M_{ij}, \ldots, M_{in}), M_{ij} = \{0,1\},$$

observed by the monitor $A_j$. Column of the table describes the relation between the functional block, test segments and monitors $M_j = B_j \approx f(T, A)$. In the monitor engine the simulated time can be introduced, which complicates the activation matrix, and indicate real or simulated cycle, on which monitoring the node or functional block states on the test segment $A_j = f(T_i, B_j, t_j)$ is performed.

At the modeling stage of fault diagnosis the response (column vector) $m = \{m_1, m_2, \ldots, m_i, \ldots, m_p\}$ of monitor engine A on the test segments T is determined, by $m_i = f(T_i, A_i)$. Searching faulty functional blocks is based on the definition of xor-operation between the assertion state vector and columns of the functional violation table $m \oplus (M_1 \vee M_2 \vee \ldots \vee M_j \vee \ldots \vee M_n)$. The choice of solution is realized by using a method for xor-analysis columns, to choose a set of vectors $B_j$ with minimum number of unit coordinates:

$$B = \min_{j=1,n} [B_j = \sum_{i=1}^{p} (B_{ij} \oplus m_i)],$$

forming the functional blocks with faults, verified on the test segments. In addition to the model for matrix diagnosis is necessary to describe the following important features of the matrix:

1) $M_i = (T_i - A_j)$;
2) $\bigvee_{i=1}^{m} M_{ij} \to \forall M_j = 1$;
3) $M_{ij} \oplus_{j=1}^{n} M_{rj} \neq M_{ij}$;
4) $M_{ij} \oplus_{i=1}^{k} M_{ir} \neq M_{ij}$;
5) $\log_2 n \leq k \leftrightarrow \log_2 |B| \leq |T|$
6) $B_j = f(T, A) \to B \oplus T \oplus A = 0.$

The features means:
1) Each row of the matrix is a match or subset of the Cartesian product (test – monitor).

2) Disjunction of all rows of the matrix gives a vector equal to one over all the coordinates.
3) All rows are distinct, which eliminates the test redundancy.
4) All columns of the matrix are distinct, which exclude the existence of equivalent faults.
5) The number of matrix rows must be greater than the binary logarithm of the columns number that determines the potential diagnosability of all blocks. 6) Diagnosis function for block depends on the complete test and monitors, which must be minimized without reduction the diagnosability.

## II. DIAGNOSABILITY OF SOC HDL-MODEL

As for the model quality for diagnosis functional violations, it shows the efficiency of the use of pair (test, assertions) for a given diagnosis depth. Evaluation of the model quality is functionally dependent on the length of the test $|T|$, a assertions number $|A|$, and detected blocks number with functional violation $N_d$ on the total number of software blocks N:

$$Q = E \times D = \frac{]\log_2 N[}{|T| \times |A|} \times \frac{N_d}{N}.$$

The diagnosis efficiency is the ratio of the minimum number of bits needed for identification (recognition) of all the blocks to the real number of code bits, presented by the product of test length by number of assertions in each of them. If the first estimate fraction is equal to 1 and all the blocks with functional violation are detected ($N_d = N$), it means a test and assertions are optimal that gives value of 1 for quality criterion of diagnosis model.

Evaluation of the structure quality for the design code is interested as the perspective of the diagnosability of software blocks. The purpose of the analysis is determining the quantitative assessment of the graph structure and a node for placement of assertion monitors, which make possible to obtain maximum diagnosis depth of functional violation of the software components. It is important not controllability and observability as in testability, but the distinguish ability of the software components with functional violation, in the limit it is zero blocks with the equivalent (indistinguishable) violation. Such an assessment may be useful to compare the graphs implemented the same functionality. It is necessary to evaluate the graph structure from the position of potential detection depth of software functional violation. One possible option is diagnosability of ABC-graph as a function depending on adjacent arcs of each node (the number $N_n$), one of which is incoming, other one is outgoing. These arcs form paths without reconvergent fan-outs and branching (N is total number of arcs in the graph):

$$D = \frac{N - N_n}{N}.$$

Each node, joining two arcs entering in the number $N_n$, is called transit one. The estimation $N_n$ is the number of indistinguishable functional violation of the software components. Potential locations of monitors for distinguishing functional violation are transit nodes. Given the above estimation of the diagnosability D the diagnosis model quality for software takes the form:

$$Q = E \times D = \frac{]\log_2 N[}{|T| \times |A|} \times \frac{N - N_n}{N}.$$

Rules for synthesis of diagnosable software:

1) Test (testbench) must create minimum number of one-dimensional activation paths, covered all the nodes and arcs of ABC-graph.

2) The base number of monitor-assertions equals the number of end nodes of the graph with not outgoing arcs.

3) An additional monitor can be placed in each node, which has one incoming and one outgoing arc.

4) Parallel independent code blocks have n monitors and a single test or one integrated monitor and n tests.

5) Serially connected blocks have one activation test for serial path and n-1 monitor or n tests and n monitors.

6) The graph nodes, which have different numbers of input and output arcs, create the conditions for the diagnosability of current section by one-dimensional activation tests without having to install additional monitors.

7) The set of test segments (testbench) has to be 100% functional coverage, given by the nodes of ABC-graph.

8) Diagnosability function is directly proportional to the test length, the number of assertions and inversely proportional to the binary logarithm of the number of software blocks:

$$D = \frac{N - N_n}{N} = f(T, A, N) = \frac{|T| \times |A|}{]\log_2 N[}.$$

Diagnosability as a function depending on the graph structure (for software), test and assertion monitors can always be reduced to unit value. For this purpose there are two alternative ways. The first one is increase of test segments, activating new paths for distinguishing equivalent faults without increasing assertions, if the software graph structure allows the potential links. The second is placement of additional assertion monitors in transit nodes of the graph. A third hybrid variant is possible, based on the joint application of two above ways. The relation of three components (the number of software blocks, the power of assertion engine and the test length) forms the set of optimal solutions

$$D = 1 \rightarrow \frac{|T| \times |A|}{]\log_2 N[} = 1 \rightarrow ]\log_2 N[ = |T| \times |A|,$$

when quality of the diagnosis and diagnosability model is equal to 1. It can be useful for choosing an quasioptimal variant of alternative way for providing the full distinguish ability of software functional violation on a pair $|T| \times |A|$.

### III. MODEL FOR DETECTING FUNCTIONAL FAILURES IN SOFTWARE

An analytic model for verification of HDL-code by using temporal assertion engine (additional observation lines) is focused to achievement the specified diagnosis depth and presented as follows:

$$M = f(F, A, B, S, T, L), \quad F = (A * B) \times S; \; S = f(T, B);$$
$$A = \{A_1, A_2, \ldots, A_i, \ldots, A_n\}; \quad B = \{B_1, B_2, \ldots, B_i, \ldots, B_n\};$$
$$S = \{S_1, S_2, \ldots, S_i, \ldots, S_m\}; \quad S_i = \{S_{i1}, S_{i2}, \ldots, S_{ij}, \ldots, S_{ip}\};$$
$$T = \{T_1, T_2, \ldots, T_i, \ldots, T_k\}; \quad L = \{L_1, L_2, \ldots, L_i, \ldots, L_n\}.$$

Here $F = (A * B) \times S$ is functionality, represented by Code-Flow Transaction Graph – CFTG (Fig. 1); $S = \{S_1, S_2, \ldots, S_i, \ldots, S_m\}$ are nodes or states of software when simulating test segments. Otherwise the graph can be considered as ABC-graph – Assertion Based Coverage Graph. Each state $S_i = \{S_{i1}, S_{i2}, \ldots, S_{ij}, \ldots, S_{ip}\}$ is determined by the values of design essential variables (Boolean, register variables, memory). The oriented graph arcs are represented by a set of software blocks

$$B = (B_1, B_2, \ldots, B_i, \ldots, B_n), \; \bigcup_{i=1}^{n} B_i = B; \; \bigcap_{i=1}^{n} B_i = \varnothing,$$

where the assertion $A_i \in A = \{A_1, A_2, \ldots, A_i, \ldots, A_n\}$ can be put in correspondence to each of them. Each arc $B_i$ – a sequence of code statements – determines the state of the node $S_i = f(T, B_i)$ depending on the test $T = \{T_1, T_2, \ldots, T_i, \ldots, T_k\}$. The assertion monitor, uniting the assertions of node incoming arcs $A(S_i) = A_{i1} \vee A_{i2} \vee \ldots \vee A_{ij} \vee \ldots \vee A_{in}$ can be put in correspondence to each node. A node can have more than one incoming (outcoming) arc. A set of functionally faulty blocks is represented by the list $L = \{L_1, L_2, \ldots, L_i, \ldots, L_n\}$.

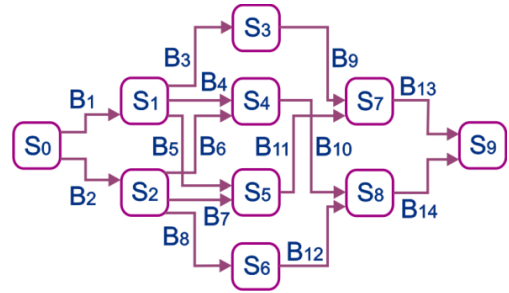

$B = (B_1 B_3 B_9 \vee (B_2 B_7 \vee B_1 B_5) B_{11}) B_{13} \vee$
$\vee ((B_1 B_4 \vee B_2 B_6) B_{10} \vee B_2 B_8 B_{12}) B_{14} =$
$= B_1 B_3 B_9 B_{13} \vee B_2 B_7 B_{11} B_{13} \vee B_1 B_5 B_{11} B_{13} \vee$
$\vee B_1 B_4 B_{10} B_{14} \vee B_2 B_6 B_{10} B_{14} \vee B_2 B_8 B_{12} B_{14}.$

Figure 1. Example of ABC-graph for HDL-code

The model for HDL-code, represented in the form of ABC-graph, describes not only software structure, but test slices of the functional coverage, generated by using software blocks, incoming to the given node. The last one defines the relation between achieved on the test variable space and potential one, which forms the functional coverage as the power of state i-th graph node $Q = \text{card} C_i^r / \text{card} C_i^p$. In the aggregate all nodes have to be full coverage of the state space of software variables, which determines the test quality, equal to 1 (100%): $Q = \text{card} \bigcup_{i=1}^{m} C_i^r / \text{card} \bigcup_{i=1}^{m} C_i^p = 1$.
Furthermore, the assertion engine $<A,C>$ that exists in the graph allows monitoring arcs (code-coverage) $A = \{A_1, A_2,...,A_i,...,A_n\}$ and nodes (functional coverage) $C = \{C_1, C_2,...,C_i,...,C_m\}$. The assertions on arcs are designed for diagnosis of the functional failures in software blocks. The assertions on graph nodes carry information about the quality of test (assertion) for their improvement or complement. The Code-Flow Transaction Graph makes possible the following: 1) use the testability design to estimate the software quality; 2) estimate the costs for creating tests, diagnosing and correcting the functional failures; 3) optimize test synthesis by means of solving the coverage problem by the minimum set of activated paths of all arcs (nodes). For instance, the minimum test for the above mentioned ABC-graph has six segments, which activate all existent paths:

$$T = S_0 S_1 S_3 S_7 S_9 \vee S_0 S_1 S_4 S_8 S_9 \vee S_0 S_1 S_5 S_7 S_9 \vee$$
$$\vee S_0 S_2 S_4 S_8 S_9 \vee S_0 S_2 S_5 S_7 S_9 \vee S_0 S_2 S_6 S_8 S_9.$$

Tests can be associated with the following program block activation matrix:

| $B_{ij}$ | $B_1$ | $B_2$ | $B_3$ | $B_4$ | $B_5$ | $B_6$ | $B_7$ | $B_8$ | $B_9$ | $B_{10}$ | $B_{11}$ | $B_{12}$ | $B_{13}$ | $B_{14}$ |
|---|---|---|---|---|---|---|---|---|---|---|---|---|---|---|
| $T_1$ | 1 | . | 1 | . | . | . | . | . | 1 | . | . | . | 1 | . |
| $T_2$ | 1 | . | . | 1 | . | . | . | . | . | 1 | . | . | . | 1 |
| $T_3$ | 1 | . | . | . | 1 | . | . | . | . | . | 1 | . | 1 | . |
| $T_4$ | . | 1 | . | . | . | 1 | . | . | . | 1 | . | . | . | 1 |
| $T_5$ | . | 1 | . | . | . | . | 1 | . | . | . | . | 1 | 1 | . |
| $T_6$ | . | 1 | . | . | . | . | . | 1 | . | . | . | 1 | . | 1 |

The activation matrix shows the fact of indistinguishability of the functional failures on a test in the blocks 3 and 9, 8 and 12, which constitute two equivalence classes if there is one assertion (monitor) in the node 9. To resolve this indistinguishability it is necessary to create two additional monitors in the nodes 3 and 6. As a result, three assertions in the nodes $A = (A_3, A_6, A_9)$ allow distinguishing all the blocks of software code. Thus, the graph enables not only to synthesize the optimal test, but also to determine the minimum number of assertion monitors in the nodes to search faulty blocks with a given diagnosis depth.

IV. MULTILEVEL METHOD (ENGINE) FOR DIAGNOSIS OF DIGITAL SYSTEM

Process model or method for searching faults by diagnosis multi-tree is reduced to creation of the engine (Fig. 2) for traversal of tree branch on the depth, specified by the user:

$$B_j^{rs} \oplus A^{rs} = \begin{cases} 0 \to \{B_j^{r+1,s}, R\}; \\ 1 \to \{B_{j+1}^{rs}, T\}. \end{cases}$$

Here vector xor-operation is executed between the columns of the matrix and the output response vector $A^{rs}$, which is determined by the functionality response taken from the monitors (assertions or bits of boundary scan register) under all test patterns. If at least one coordinate of vector xor-sum is equal to zero $B_j^{rs} \oplus A^{rs} = 0$ then one of the following action is performed: the transition to the activation matrix of lower level $B_j^{r+1,s}$ or repair of the functional block $B_j^{rs}$. At that analysis is carried out, what is the most important: 1) time – then repair of faulty block is performed; 2) money – then a transition down is carried out to specify the fault location, because replacement of smaller block substantially decreases the repair cost. If at least one coordinate of the resulting xor-sum vector is equal to one $B_j^{rs} \oplus A^{rs} = 1$, then transition to the next matrix column is performed. When all coordinates of the assertion monitor vector are equal to zero $A^{rs} = 0$, fault-free state of a device is fixed. If all vector sums are not equal to zero $B_j^{rs} \oplus A^{rs} = 1$, it means a test, generated for check the given functionality, has to be corrected.

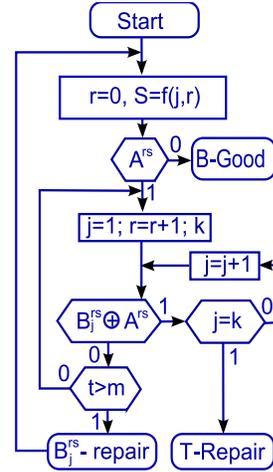

Figure 2. Engine for traversal of diagnosis multitree

Thus, the dataflow shown in the Fig. 2, allows realizing efficient infrastructure IP for complex technical systems. The advantages of the engine, which is invariant to the hierarchy levels, are the simplicity of preparation and presentation of diagnostic information in the form of minimizing activation table of functional blocks on the test patterns.

V. VERIFICATION OF MODELS AND METHOD FOR DIAGNOSIS

To illustrate the performance of the proposed model and method the functionalities of three modules of the digital filter of Daubechies [11] are considered below. The first component is component Row_buffer; its transaction graph, based on RTL-model, shown in Fig. 3. Nodes are presented by the states of variables and monitors, which are responsible for node incoming transactions or arcs, corresponding to the functional blocks.

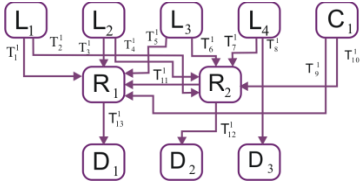

Figure 3. Component Row_buffer of transaction graph

An activation table for functional blocks is generated by using graph, obtained during the simulation (Table I). Table rows are activation paths for blocks to the given monitor-node. A table is a coverage all columns or functional blocks by rows of paths. In this case it should not have at least two identical columns. The difference of table is creation of the pair <test – observed node>, making it possible to considerably reduce the dimension of the table with 100% detection of all faulty blocks. The main feature of the proposed model is the ability to describe the following relations by using the table: distinct tests – one node, one test – distinct nodes.

TABLE I. ACTIVATION TABLE

| $A_{ij}$ | $T_1$ | $T_2$ | $T_3$ | $T_4$ | $T_5$ | $T_6$ | $T_7$ | $T_8$ | $T_9$ | $T_{10}$ | $T_{11}$ | $T_{12}$ | $T_{13}$ |
|---|---|---|---|---|---|---|---|---|---|---|---|---|---|
| $t_1 \to D_3$ | . | . | . | . | . | . | . | 1 | . | . | . | . | . |
| $t_2 \to D_1$ | 1 | . | . | . | . | . | . | . | . | . | . | . | 1 |
| $t_3 \to D_1$ | . | . | 1 | . | . | . | . | . | . | . | . | . | 1 |
| $t_4 \to D_1$ | . | . | . | . | 1 | . | . | . | . | . | . | . | 1 |
| $t_5 \to D_1$ | . | . | . | . | . | . | . | . | . | . | 1 | . | 1 |
| $t_6 \to D_1$ | . | . | . | . | . | . | . | . | 1 | . | . | . | 1 |
| $t_7 \to D_2$ | . | 1 | . | . | . | . | . | . | . | . | . | 1 | . |
| $t_8 \to D_2$ | . | . | . | 1 | . | . | . | . | . | . | . | 1 | . |
| $t_9 \to D_2$ | . | . | . | . | . | 1 | . | . | . | . | . | 1 | . |
| $t_{10} \to D_2$ | . | . | . | . | . | . | 1 | . | . | . | . | 1 | . |
| $t_{11} \to D_2$ | . | . | . | . | . | . | . | . | . | 1 | . | 1 | . |

The use of the activation matrix of functional blocks (transaction graph) and xor-method for detecting faults allows synthesizing logic functions for the combination circuit, which determines number of functional block with semantic errors in process of simulation:

$$D_3 = T_8^1;$$

$$D_1 = T_{13}^1 T_1^1 \vee T_{13}^1 T_3^1 \vee T_{13}^1 T_5^1 \vee T_{13}^1 T_{11}^1 \vee T_{13}^1 T_9^1;$$

$$D_2 = T_{12}^1 T_2^1 \vee T_{12}^1 T_4^1 \vee T_{12}^1 T_6^1 \vee T_{12}^1 T_7^1 \vee T_{12}^1 T_{10}^1.$$

This feature is possible due to the lack of equivalent faults or identical columns in the activation matrix. Therefore, fixing the actual state of monitors at the nodes $D_1, D_2, D_3$ on 11 test patterns makes it possible to unambiguously identify an incorrect functional module by performing xor-operation between the assertion vector and the columns of activation matrix. Zero value of all coordinates for the result of xor-operations determines the number of the column corresponding to a faulty module. The implementation of the model and method in a logical function allows identifying the faulty block before the completion of the diagnosing experiment, if it is possible. This means significant savings of diagnosis time for certain types of faults. For instance, the test-monitor $t_1 \to D_3$ allows identifying a fault of the block $B_8$ at the first test.

Second test case for the practical use of the activation model and xor-method for searching faults is presented below. Synthesis of the diagnosis matrix for discrete cosine transform module from Xilinx library in the form of functional coverage is shown in Listing 1.

Listing 1. Part of functional coverage
```
c0: coverpoint xin
{
bins minus_big={[128:235]};
bins minus_sm={[236:255]};
bins plus_big={[21:127]};
bins plus_sm={[1:20]};
bins zero={0};
}
c1: coverpoint dct_2d
{
bins minus_big={[128:235]};
bins minus_sm={[236:255]};
bins plus_big={[21:127]};
bins plus_sm={[1:20]};
bins zero={0};
bins zero2=(0=>0);
}
endgroup
```

For all 12 modules the transaction graphs, activation tables, and logic functions are developed for testing and fault detection in the discrete cosine transform. Graph with the activation matrix and logic function (Fig. 4) are presented below.

This graph is associated with the following diagnosis matrix (Table II).

TABLE II. DIAGNOSIS MATRIX

| $A_{ij}$ | $T_1$ | $T_2$ | $T_3$ | $T_4$ | $T_5$ | $T_6$ | $T_7$ | $T_8$ | $T_9$ | $T_{10}$ | $T_{11}$ | $T_{12}$ | $T_{13}$ | $T_{14}$ |
|---|---|---|---|---|---|---|---|---|---|---|---|---|---|---|
| $P_1 \to F_7$ | 1 | . | 1 | . | 1 | . | 1 | . | . | . | . | . | . | . |
| $P_2 \to F_8$ | . | 1 | . | 1 | 1 | . | . | 1 | . | . | . | . | . | . |
| $P_3 \to F_9$ | 1 | . | 1 | . | . | 1 | . | . | . | . | 1 | . | . | . |
| $P_4 \to F_{10}$ | . | 1 | . | 1 | . | 1 | . | . | . | . | . | 1 | . | . |
| $P_5 \to F_{12}$ | 1 | . | 1 | . | 1 | . | . | . | 1 | . | . | . | 1 | . |
| $P_6 \to F_{13}$ | . | 1 | . | 1 | . | 1 | . | . | . | 1 | . | . | . | 1 |
| $P_1 \to F_2$ | 1 | . | . | . | . | . | . | . | . | . | . | . | . | . |
| $P_2 \to F_3$ | . | 1 | . | . | . | . | . | . | . | . | . | . | . | . |

The system of diagnosis functions is presented below:

$$F_7 = T_1^1 T_3^1 T_5^1 T_7^1; \quad F_8 = T_2^1 T_4^1 T_5^1 T_8^1; \quad F_9 = T_{11}^1 T_6^1 T_1^1 T_3^1;$$

$$F_{10} = T_4^1 T_5^1 T_6^1 T_{12}^1; \quad F_{12} = T_1^1 T_3^1 T_5^1 T_9^1 T_{13}^1;$$

$$F_{13} = T_2^1 T_4^1 T_6^1 T_{10}^1 T_{14}^1;$$

$$F_2 = T_1^1; \quad F_3 = T_2^1;$$

Fragment of monitor engine is presented by Listing 2.
Listing 2. Code fragment of monitor engine
```
sequence first( reg[7:0] a, reg[7:0]b);
 reg[7:0] d;
 (!RST,d=a)
##7 (b==d);
 endsequence
property f(a,b);
 @(posedge CLK)
 // disable iff(RST||$isunknown(a)) first(a,b);
 !RST |=> first(a,b);
endproperty
```

```
odin:assert property (f(xin,xa7_in))
 // $display("Very good");
else $error("The  end,  xin  =%b,xa7_in=%b",  $past(xin,
7),xa7_in);
```

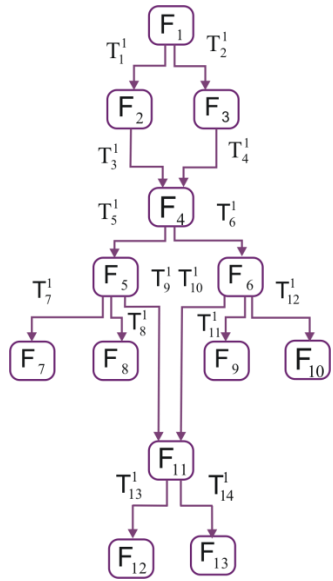

Figure 4.  Transaction graph of main-RTL module

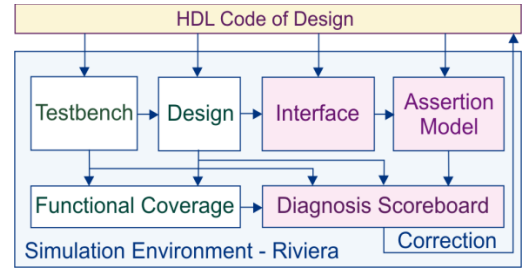

Figure 5.  Implementation of results in the system Riviera

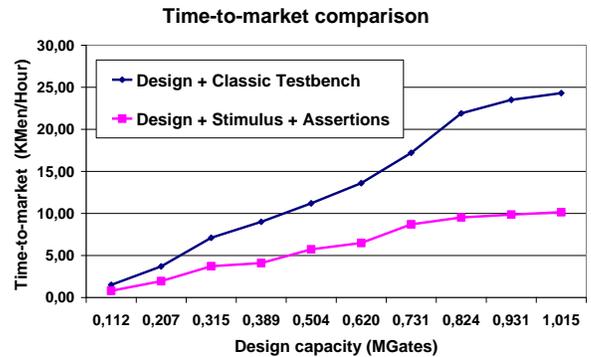

Figure 6.  Comparative analysis of verification methods

Testing of discrete cosine transformation in the environment Riviera, Aldec detects incorrectness in seven rows of HDL-models:

//add_sub1a <= xa7_reg + xa0_reg;//

Subsequent correcting code allowed obtaining the following code (Listing 3).

Listing 3. Corrected code fragment

```
add_sub1a <= ({xa7_reg[8],xa7_reg} + {xa0_reg[8],xa0_reg});
add_sub2a <= ({xa6_reg[8],xa6_reg} +{xa1_reg[8],xa1_reg});
add_sub3a <= ({xa5_reg[8],xa5_reg} +{xa2_reg[8],xa2_reg});
add_sub4a <= ({xa4_reg[8],xa4_reg} + {xa3_reg[8],xa3_reg});
end
else if (toggleA == 1'b0)
begin
add_sub1a <= ({xa7_reg[8],xa7_reg} - {xa0_reg[8],xa0_reg});
add_sub2a <= ({xa6_reg[8],xa6_reg} - {xa1_reg[8],xa1_reg});
add_sub3a <= ({xa5_reg[8],xa5_reg} - {xa2_reg[8],xa2_reg});
add_sub4a <= ({xa4_reg[8],xa4_reg} - {xa3_reg[8],xa3_reg});
```

## VI. IMPLEMENTATION OF MODELS AND METHODS IN THE VERIFICATION SYSTEM

Practical implementation of models and verification methods is integrated into the simulation environment Riviera of Aldec Inc., Fig. 5. New assertion and diagnosis modules, added in the system, improved the existing verification process, which allowed 15% reduction the design time of digital product.

Actually, application of assertions makes possible to decrease the length of test-bench code and considerably reduce (x3) the design time (Fig. 6), which is the most expensive. Assertion engine allows increasing the diagnosis depth of functional failures in software blocks up to level 10-20 HDL-code statements.

Due to the interaction of simulation tools and assertion engine, automatically placed inside the HDL-code, an access of diagnosis tools to the values of all internal signals is appeared. This allows quickly identifying the location and type of the functional violation, as well as reducing the time of error detection in the evolution of product with top-down design. Application of assertion for 15 real-life designs (from 5 thousand up to 5 million gates) allowed obtaining hundreds of dedicated solutions, included in the verification template library VTL, which generalizes the most popular on the EDA market temporal verification limitations for the broad class of digital products. Software implementation of the proposed system for analyzing assertions and diagnosis HDL-code is part of a multifunctional integrated environment Aldec Riviera for simulation and verification of HDL-models.

High performance and technological combination of assertion analysis system and HDL-simulator of Aldec Company is largely achieved through integration with the internal simulator components, including HDL-language compilers. Processing the results of the assertion analysis system is provided by a set of visual tools of Riviera environment to facilitate the diagnosis and removal of functional violation. The assertion analysis model can also be implemented in hardware with certain constraints on a subset of the supported language structures. Products Riviera including the components of assertion temporal verification, which allow improving the design quality for 3-5%, currently, occupies a leading position in the world IT market with the number of installations of 5,000 a year in 200 companies and universities in more than 20 countries on the world.

## VII. Conclusion

1. Proposed transactional graph model and method for diagnosis of digital systems-on-chips are focused to considerable reducing the time of functional violation based on diagnosis matrix with ternary relations in the form of test, monitor, and functional component.

2. An improved process model for detection of functional violation in software or hardware is proposed. It is characterized by using the xor-operation, which makes it possible to improve the diagnosis performance for single and multiple faults (functional violation) on the basis of parallel analysis of the fault table, boundary scan standard IEEE 1500, and vector operations and, or, xor.

3. A model for diagnosis the functionality of digital system-on-chip in the form of multi-tree and method for tree traversal, implemented in the engine for faults detection with given depth, are developed. They considerably increase the performance of software and hardware Infrastructure IP.

4. Test verification of proposed diagnosis method is performed by three real case studies, presented by SoC functionalities of a cosine transform filter, which showed the consistency of the results in order to minimize the time of fault detection and memory for storing diagnosing information, as well as increase the diagnosis depth for digital products.